\documentclass[11pt]{article}
\usepackage{mathrsfs}
\usepackage{esdiff}
\usepackage{amsthm,amsfonts}
\usepackage{amsmath}
\usepackage{amssymb}
\usepackage{graphics,epsfig}
\usepackage{stackrel}

\newtheorem{lemma}{Lemma}[section]

\newcommand{\beq}{\begin{equation}}
\newcommand{\enq}{\end{equation}}

\newcommand{\beqa}{\begin{eqnarray}}
\newcommand{\enqa}{\end{eqnarray}}
\newcommand{\beqas}{\begin{eqnarray*}}
\newcommand{\enqas}{\end{eqnarray*}}

\newcommand{\bep}{\par\noindent {\sc Proof. }}
\newcommand{\enp}{$\bigtriangleup$\smallskip\par}

\newcommand{\E}{ {\bf E}}

\newtheorem{thm}{Theorem}[section]

\newtheorem{prop}{Proposition}[section]

\numberwithin{equation}{section}

\begin{document}
\author{
Elena Boguslavskaya\thanks{elena@boguslavsky.net}\,\,\,\thanks{The author is supported by Daphne Jackson fellowship funded by ESPRC} \\
Brunel University London, United Kingdom, \and \\
Dmitry Muravey\thanks{dmuravey@hse.ru}\\
Internatonal Laboratory of Quantitative Finance, \\ Higher School of Economics, Moscow, Russia
}
\date{}

\title
{
     An explicit solution for optimal investment in Heston model\thanks{The research on this paper was supported by grant 14.12.31.2007 of the Russian government.}
}

\maketitle
\abstract
{
    In this paper we consider a variation of the Merton's problem with added stochastic volatility and finite time horizon. It is known that the corresponding optimal control problem may be reduced to a linear parabolic boundary problem under some assumptions on the underlying process and the utility function. The resulting parabolic PDE is often quite difficult to solve, even when it is linear. The present paper contributes to the pool of explicit solutions for stochastic optimal control problems. Our main result is the exact solution for optimal investment  in Heston model.
}

\section{Introduction}

The optimal control theory for stochastic processes plays a prominent role in financial mathematics
 and allows to formulate and solve problems in optimal investment, optimal trading, and other areas related to finance. The seminal paper by Robert Merton \cite{Merton} was generalized in several directions. The subsequent research has introduced more realistic and complex asset dynamics
\cite{Zar01}, \cite{Zar02}, and more realistic market models
\cite{Kabanov}. Similar mathematical problems arise in algorithmic trading
\cite{Boguslavskaya}, and market microstructure studies \cite{Obizhaeva}.

In this paper we consider a variation of the Merton's problem with stochastic volatility and finite time horizon. As shown in \cite{Zar01}, the optimal control problem may be reduced to a linear parabolic boundary problem under some assumptions on the underlying process and the utility function.
 Generally, due to the lack of solution smoothness for this boundary  problem, one needs to find a viscosity solution to a non-linear equation.

The resulting parabolic PDE is often quite difficult to solve, even when it is linear. In several special cases the explicit solutions were obtained, see \cite{Kraft} for the model similar to Heston's, and \cite{Chacko} for Chacko-Viceira model.

The present paper contributes to the pool of explicit solutions for stochastic optimal control problems.
Our main result is the exact solution to the optimal control problem within the framework of the celebrated Heston model
\cite{Heston}. The method of affine representations, as used in
\cite{Kraft}, cannot be applied here. To obtain the desired explicit solution we employ the theory of integral representations and special functions. Quasi-analytical solutions were obtained by asymptotic and perturbation methods
in \cite{Zar02} for optimal control of more general processes.

The paper is organized as follows: in section \ref{sec Statement} we formulate the problem; in section \ref{sec Zar} we discuss the results of \cite{Zar01}
on the representation of Bellman function as the solution of the linear parabolic equation, and extend their results to the case of exponential utility;
  in section \ref{sec Heston} we present the main results of the paper; in section \ref{sec Analysis} we analyze the obtained explicit solution; in Appendix (section \ref{appendix}) we provide some technical asymptotics relations.

\section{Formulation of the problem}
\label{sec Statement}
 Let $(X,V) =(X_s,V_s)_{s\geq t}$ be a vector stochastic process, given by the triangular system of stochastic differential equations
\beqa
        \label{model Heston}
        d X / X &=&\mu \,dt + \sqrt{V} dB^1,       \qquad   \qquad X_t = x            \\
        d V &=& k (\Theta - V)ds + \sigma \sqrt{V} d B^2,   \qquad V_t = v. \nonumber
    \enqa
where $B^1 = (B^1_t)_{s \geq t}$ è $B^2 = (B^2_t)_{s \geq t}$
are correlated Wiener processes with correlation coefficient
 $\rho$, i.e.  $\langle B^1, B^2\rangle_t = \rho \, t$, and $\mu$, $k$, $\Theta$, $\sigma$ are some constants. The model for the asset price $X$ defined by (\ref{model Heston}) is very well known in financial mathematics, and is called the Heston model \cite{Heston}.

 In order to solve the system of SDE's (\ref{model Heston}) one needs to solve the second equation first to obtain $V$, and then solve for $X$.
  Process $V=(V_s)_{s\geq t}$ represents the volatility of asset $X$. Process $V$ is a Feller process, also called a CIR (Cox-Ingersoll-Ross) process in financial mathematics \cite{Cox} .
   We assume that the Feller condition $2k\Theta > \sigma^2$ is satisfied for $V$. Thus,  $V$ is always positive given $V_0>0$.

   Let us consider a controlled process $W = (W_s)_{s \geq t}$ given by
    \beq
        \label{model W}
        dW = \alpha dX, \qquad W_t =w.
    \enq
where control $\alpha=(\alpha_s)_{s \geq t}$ is adapted to filtration ${\cal F}_{\geq t} =\sigma \{B_s -B_t, \, s \geq t \}$ and $\int_t^T \alpha_s^2 dt <\infty$.

Our aim is to find  the control on which functional $\E  U( W_T)$ achieves its maximum
    \beq
        \label{model J}
        J(w,x,v,t) = \sup_{\alpha} \E  U( W_T).
    \enq
 Function $J(w,x,v,t)$ is the Bellman function, which is a solution to HJB ( Hamilton-Jacobi-Bellman) equation. Utility function $U(w)$ can be a power or an exponential utility function.
 For the convenience of notation we will omit the indices.
    \beq
        \label{model U}
        U_P(w) = \frac{w^\gamma}{\gamma} , \quad \gamma<0, \qquad\qquad U_E(w) = 1-\frac{e^{-c w}}{c},  \quad c > 0.
    \enq
Note, that the logarithmic utility is a special case of the power utility with $\gamma=0$. This well known observation can be checked by taking the limit
$\lim_{\gamma \rightarrow 0} (w^{\gamma} - 1)/ \gamma = \log(w) $ and noting that utility functions are defined up to an additive constant.

The financial interpretation of the processes above is the following: process $W$ represents the wealth dynamics of the investor, while control $\alpha$ is the evolution of the investor's position in the asset. The utility function $U$ represents the investor's preferences. Thus, to invest optimally,  one should find the optimal investment rule at each time $t$ such that the expected utility of the terminal wealth at  time $T$  is maximised (\ref{model J}).

Function $J(w,x,v,t)$ is the solution of Hamilton-Jacobi-Bellman (HJB) equaion. The solution of HJB equation in the general case is a non-linear PDE. However, this equation may become linear under some choice of utility functions.
Zariphopoulou \cite{Zar01} has proposed a method based on viscosity solution technique. Under this method the equation
(\ref{model J}) may be reduced to a linear parabolic equation. The method may be applied to systems of stochastic differential equations of ``triangular type''. Under this restriction the drift and the diffusion of the underlying asset $X$ may depend on the volatility process $V$ in arbitrary way.

It is much easier to perform qualitative and quantitative analysis of the optimal solution while provided with the exact formula.
Nevertheless, it is not always possible to find an explicit solution to a linear parabolic PDE.  Explicit solutions were obtained in some special cases: see [7] for the model similar to Heston's, and [8] for Chacko-Viceira model. Either of the above results are based on affine representations of solutions. This technique is not applicable in the model we consider  (\ref{model Heston}).

In the present paper we propose a method based on Laplace transform. This technique allows us  to obtain an explicit solution in terms of confluent hypergeometric functions.
\par

\section{T. Zariphopoulou's result and extensions}
\label{sec Zar}
     A quite general model for optimal investment (\ref{model J})  with power utility function  $U_P$ was considered in \cite{Zar01}. The asset price dynamics $X$ dependent on random process $V$ was given by the triangular system of stochastic differential equations
    \beqa
        \label{zar process}
        dX / X &=&  \mu (V,t) dt + \sigma(V,t) dB^1,\nonumber\\
        dV &=& b(V,t)dt + a(V,t) d B^2,
    \enqa
    where $B^1 = (B^1_t)_{s \geq t}$ è $B^2 = (B^2_t)_{s \geq t}$
are correlated Wiener processes with correlation coefficient
 $\rho$.

    We assume the necessary restrictions on the coefficients of system (\ref{zar process}) being satisfied in order to guarantee the existence of a strong solution $(X,V)$.

    \begin{thm}[Zariphopoulou]
        \label{zar thm power}
    For process (\ref{zar process}) and power utility function (\ref{model U})
    we have the following representation of Bellman function  (\ref{model J})
        \beq
            \label{zar J}
            J_P(w,x,v,t) = \frac{w^\gamma}{\gamma} f^{1/\delta} (v,t),
        \enq
        where
        \beq
        \label{delta power}
        \delta = 1+\rho^2\frac{\gamma }{1-\gamma},
        \enq
        and function $f=(v,t)$ is a solution to the following parabolic boundary value problem
        \beqa
            \label{zar f}
                f_t + \frac{1}{2} a^2(v,t)\, f_{vv} +
                    \left(b(v,t) +\rho\frac{\gamma \, \mu(v,t) a(v,t)}{(1-\gamma)\sigma(v,t)}\right) f_v +
                    \frac{\gamma}{1-\gamma} \frac{\delta}{2} \frac{\mu^2(v,t)}{\sigma^2(v,t)}\, f =0, \\ \nonumber
                f(v,T) = 1.
        \enqa
        The optimal synthesis is given by
        \beq
            \alpha_{P}^{*} (w, x, v, t) = \frac{w}{x (1-\gamma)} \left( \frac{\mu(v,t)}{\sigma^2(v,t)} + \frac{\rho} {\delta} \frac{a(v,t) }{\sigma(v,t)}\frac{f_v (v,t)}{ f(v,t)} \right).
            \label{zar synt}
        \enq
    \end{thm}

    To find the optimal control at time $t$ one should substitute the values for processes $W$ and $(X,V)$ into synthesis
    (\ref{zar synt}). Note, that representation (\ref{zar J})
    can be explained as follows. Firstly, one can separate variables in the Bellman function $J(w,x,v,t) = u(w)R(x,v,t)$. This trick is well known for the power utility case. Secondly, from the form of the coefficients and the boundary condition, one can see that the solution does not depend on variable $x$. Thirdly, after the above manipulations, the remaining non-linear equation can be linearised by substitution
    $R^{\delta}(v,t) = f(v,t)$.

    We use similar reasoning as above to extend Zariphopoulou's formula to the case of exponential utility.  It is our first result in this paper.
    \begin{prop}
    \label{zar prop} Under exponential utility function $U_E$,
        the Bellman function  (\ref{model J}) is represented by
        \beq
            \label{zar J exp}
            J_E(x,w,v,t) = 1-\frac{e^{-c w}}{c} f^{1/{\delta}} (v,t),
        \enq
        where
        \beq
        \label{delta exp}
         \delta = 1-\rho^2,
        \enq
        and  function $f=f(v,t)$ is a solution to the following parabolic boundary value problem
        \beqa
            \label{zar f exp}
                f_t + \frac{1}{2} a^2(v,t) f_{vv} +
                    \left(b(v,t) -\rho\frac{\mu(v,t) a(v,t)}{\sigma(v,t)}\right) f_v -
                    \frac{\delta}{2}\frac{ \mu^2(v,t)}{\sigma^2(v,t)} f = 0, \\ \nonumber
                f(v,T) = 1.
        \enqa

        The optimal synthesis is given by
        \beq
            \alpha_{E}^{*} (w, x, v, t) = \frac{1}{c \,x} \left( \frac{\mu(v,t)}{\sigma^2(v,t)} +
                \frac{\rho} {\delta} \frac{a(v,t)} {\sigma(v,t)}\frac{f_v (v,t)}{ f(v,t)} \right).
            \label{zar control}
        \enq
    \end{prop}

    Note, that constant $\delta$ given by (\ref{delta exp}) and equation (\ref{zar f exp}) for $f$ can be obtained from power utility formula
     (\ref{delta power}) and  (\ref{zar f}) by taking limit  $\gamma \rightarrow -\infty$.

\section{The main result}
\label{sec Heston}
Let us return to the original problem. Consider the Heston model given by (\ref{model Heston}).
      Using the results from section \ref{sec Zar}, in particular substitutions (\ref{zar J}) and (\ref{zar J exp}) from Theorem \ref{zar thm power} and Proposition \ref{zar prop} respectively, we can reduce the original boundary value problem for Bellman function to the  following boundary value problem for function   $f=f(v,t)$
    \beqa
    \label{heston f}
            \frac{\sigma^2 v}{2} f_{vv} +
                \left( k \Theta - \frac{1-\delta}{\rho} \mu \sigma  - k v \right)f_v
                - \frac{C}{v} f + f_{t} =0. \\
            f(v,T) = 1, \nonumber
    \enqa
    Equation (\ref{heston f}) is the same for both cases of power and exponential utility functions, with the only difference in the values of constants $\delta$ and $C$.
    In the case of power utility function
    \beq
    \label{C power}
     \delta = 1+\rho^2\frac{\gamma }{1-\gamma}, \qquad \qquad    C = -\frac{\gamma}{1-\gamma}  \frac{\mu^2}{2} \delta.
    \enq
    For the exponential utility case
    \beq
    \label{C exp}
     \delta = 1-\rho^2, \qquad \qquad    C =  \frac{\mu^2}{2} \delta.
    \enq
    \begin{lemma}
         The solution of (\ref{heston f}) is given by
        \beq
        \label{answer f}
        f(v,t)  = \frac{\Gamma (\eta - \lambda + 1/2)}{ \Gamma (2\eta+1)} e^{-\Psi(v,t)/2}
                \left( \Psi(v,t) \right) ^{\lambda} M_{\lambda, \eta} \left( \Psi(v,t)\right),
        \enq
        where
        \beq
        \label{Psi}
            \Psi(v,t) = \frac{2kv}{\sigma^2 \left(e^{k(T-t)} - 1 \right)},
            \quad \lambda = -\frac{k\Theta}{\sigma^2} +\frac{(1-\delta) \mu}{\rho \sigma},
            \quad \eta = \sqrt{\left( \lambda+\frac{1}{2}\right)^2 + \frac{2 C }{\sigma^2}},
        \enq
          $M_{\lambda, \eta}(z)$ is a Whittaker's function, and $\Gamma(z)$ is a Gamma function.
          \\
          Moreover, we have
         \beq
         \label{control Heston}
        \frac{f_v}{f} = \frac{(\eta+ \lambda +1/2) }{v}
        \frac{M_{1+\lambda, \eta}\left(\Psi(v,t) \right)}{M_{\lambda, \eta} \left( \Psi(v,t) \right)}.
    \enq
    \end{lemma}
\bep \, Making substitutions
\beq
\label{change of variables}
\tilde{v}=\frac{2kv}{\sigma^2}, \quad\quad k(T-t) = \tilde{\tau}, \quad\quad f(v,t) = e^{-\lambda \tilde{\tau}}{\tilde{v}}^{\lambda}e^{\tilde{v}/2} h(\tilde{v}, \tilde{\tau})
\enq
we get the following boundary value problem for $h$
\beqa
h_{\tilde{v}\tilde{v}} + \left(-\frac{1}{4}+\frac{1/4-\eta^2}{\tilde{v}^2}\right)h = \frac{1}{\tilde{v}}h_{\tilde{\tau}}. \nonumber \\
h(\tilde{v},0) = {\tilde{v}}^{-\lambda}e^{-\tilde{v}/2}. \nonumber
\enqa
Let $G(\tilde{v};\zeta)$ be the Laplace transform of function $h(\tilde{v},\tilde{\tau})$ with respect to $\tilde{\tau}$
\[
G(\tilde{v}; \zeta) = \int_{0}^{\infty} e^{\zeta \tilde{\tau}} h(\tilde{v}, \tilde{\tau}) d\tilde{\tau}.
\]
 By denoting $\chi(\tilde{v})  = \tilde{v}^{-1-\lambda} e^{-\tilde{v}/2}$, one can see that
\beq
\label{whittaker eq}
G'' + \left(-\frac{1}{4}-\frac{\zeta}{\tilde{v}} + \frac{1/4-\eta^2}{\tilde{v}^2}\right)G = - \chi(\tilde{v}). \\
\enq
If made homogeneous, equation (\ref{whittaker eq}) is a  Whittaker equation. Whittaker functions $M_{-\zeta, \eta}(\tilde{v})$ and $W_{-\zeta, \eta}(\tilde{v})$ are two linearly independent solutions of a Whittaker equation.
One can check by substitution that the formula below is the solution of equation (\ref{whittaker eq})
\beq
\label{g}
G(\tilde{v}; \zeta) = \frac{\Gamma(1/2+\zeta+\eta)}{\Gamma(1+2\eta)}
\left( M_{-\zeta,\eta}(\tilde{v}) \int_{0}^{\tilde{v}} \chi(\varphi) W_{-\zeta,\eta}(\varphi) d\varphi
+W_{-\zeta,\eta}(\tilde{v}) \int_{\tilde{v}}^{\infty} \chi(\varphi) M_{-\zeta,\eta}(\varphi) d\varphi  \right).
\enq
To proceed further we use formula $6.669.4$ from \cite{Gradshteyn}
\beqa
\int_0^{\infty} e^{-\frac{1}{2} (a_1+a_2) t \cosh x} \coth ^{2\nu} \left( \frac{1}{2} x \right) I_{2\mu} \left(t \sqrt{a_1 a_2 } \sinh{x}\right) dx =
\frac{\Gamma \left( \frac{1}{2} +\mu - \nu \right)}{ t \sqrt{a_1 a_2} \Gamma(1+2\mu)} W_{\nu,\mu} (a_1 t) M_{\nu,\mu} (a_2 t), \nonumber\\
{\rm Re \,}  \left( \frac{1}{2} + \mu - \nu \right) > 0, \quad {\rm Re \,} \mu > 0, \quad a_1 >a_2.  \nonumber
\enqa
The above formula from \cite{Gradshteyn} allows us to rewrite (\ref{g}) as
\beq
G(\tilde{v};\zeta) = \sqrt{\tilde{v}} \int_{0}^{\infty}\int_{0}^{\infty} \varphi^{-1/2-\lambda}
e^{-\frac{\varphi}{2} - \frac{\tilde{v} + \varphi}{2}\cosh{\psi}} \tanh^{2\zeta} \left( \frac{\psi}{2}\right)
I_{2\eta}\left( \sqrt{\tilde{v} \varphi} \sinh{\psi}\right) d\varphi d\psi.
\label{G}
\enq
Applying $6.643.2$ from \cite{Gradshteyn}
\beq
\int_0^{\infty} x^{\mu -\frac{1}{2}}  e^{-\alpha x}  I_{2\nu} \left( 2 \beta \sqrt{x} \right) dx =
\frac{\Gamma \left( \mu + \nu  + \frac{1}{2}  \right)}{ \Gamma(2\nu+1)} \beta^{-1} e^{\frac{\beta^2}{2\alpha}} \alpha^{-\mu} M_{-\mu,\nu} \left( \frac{\beta^2}{\alpha }\right), \quad
{\rm Re \,}  \left( \mu + \nu +  \frac{1}{2} \right) > 0
\enq
to the internal integral in (\ref{G}), we obtain
\beq
\label{*}
G(\tilde{v};\zeta)=e^{-\tilde{v}/2} \frac{\Gamma(\eta-\lambda + 1/2)}{\Gamma(1+2\eta)} \int_{0}^{\infty}
e^{(1-\cosh\psi)\tilde{v} / 4}
\tanh^{2\zeta} \left( \frac{\psi}{2}\right)
\left( \frac{\cosh\psi + 1}{2} \right)^{\lambda}
M_{\lambda,\eta} \left(\frac{ \cosh{\psi}-1}{2} \tilde{v} \right)\frac{2d\psi}{\sinh{\psi}}.
\enq
By changing the variable of integration $\psi$  in (\ref{*}) as
\[
2\log \left[ \tanh\left( \frac{\psi}{2}\right)\right] = \nu, \quad \frac{2d\psi}{\sinh\psi} = d\nu,
\quad \cosh\psi - 1 = \frac{2e^\nu}{1-e^{\nu}},
\]
we get
\[
G(\tilde{v}; \zeta)=e^{-\tilde{v}/2} \frac{\Gamma(\eta-\lambda + 1/2)}{\Gamma(1+2\eta)} \int_{-\infty}^{0}
\exp\left\{-\frac{\tilde{v}}{2}\frac{e^\nu}{e^\nu -1} +\zeta \nu \right\}
\left( \frac{1}{e^\nu-1} \right)^{\lambda}
M_{\lambda,\eta} \left(\tilde{v}\frac{ e^\nu}{e^\nu-1} \right) d\nu.
\]
Inverting the Laplace transform, and using substitutions (\ref{change of variables}), we recover the formula for  $f(v,t)$
\beq
\label{f 1}
f(v,t) = \frac{1}{2\pi i}\frac{\Gamma(\eta-\lambda + 1/2)}{\Gamma(1+2\eta)} \int_{N-i\infty}^{N+i\infty} \int_{-\infty}^{0}
\exp\left\{-\frac{\tilde{v}}{2}\frac{e^\nu}{e^\nu -1}  \right\} e^{\zeta (\nu+\tilde{\tau})}
\left( \tilde{v} \frac{e^{-\tilde{\tau}}}{e^\nu-1} \right)^{\lambda}
M_{\lambda,\eta} \left(\tilde{v}\frac{ e^\nu}{e^\nu-1} \right) d\zeta d\nu,
\enq
 where  $N$ is a number such that all residues of the integrand  are to the right of it.

Using the well-known representation of Dirac function
\[
\frac{1}{2\pi i} \int_{N-i\infty}^{N+i\infty} e^{z \zeta} d\zeta = \delta(z),
\]
and changing the order of integration in  (\ref{f 1}), we get
\beq
\label{fvt}
f(v,t) = \frac{\Gamma(\eta-\lambda + 1/2)}{\Gamma(1+2\eta)} \int_{-\infty}^{0} \delta(\nu+\tilde{\tau})
\exp\left\{-\frac{\tilde{v}}{2}\frac{e^\nu}{e^\nu -1}  \right\}
\left( \tilde{v} \frac{e^{-\tilde{\tau}}}{e^\nu-1} \right)^{\lambda}
M_{\lambda,\eta} \left(\tilde{v}\frac{ e^\nu}{e^\nu-1} \right) d\nu.
\enq
 Note, that $\tilde{\tau} \geq 0$. Thus, we can complement the range of integration in (\ref{fvt}) to the whole line, and, using the definition of Dirac's function, namely  $\int_{-\infty}^{\infty} \delta(\zeta -z) g(\zeta)d\zeta = g(z)$ for any continuous $g$, we get
 \beq
\label{fvtt}
f(v,t) = \frac{\Gamma(\eta-\lambda + 1/2)}{\Gamma(1+2\eta)}
\exp \left\{-\frac{\tilde{v}}{2}\frac{e^{\tilde{\tau}}}{e^{\tilde{\tau}} -1}  \right\}
\left( \tilde{v} \frac{e^{-\tilde{\tau}}}{e^{\tilde{\tau}}-1} \right)^{\lambda}
M_{\lambda,\eta} \left(\tilde{v}\frac{ e^{\tilde{\tau}}}{e^{\tilde{\tau}}-1} \right).
\enq

Finally, by reversing the change of variables (\ref{change of variables}), we receive the main formula (\ref{answer f}). The expression (\ref{control Heston}) for $ f_v /f $ is obtained by using differential rules for Whittaker functions (see \cite{Abramowitz}).
\enp
\bigskip
The following theorem is the analogue of  the results from section \ref{sec Zar} but for Heston's model. It gives an exact solution for the optimal control and Bellman function.
\begin{thm}
        \label{heston thm power}
    For power utility (\ref{model U}) and process (\ref{model Heston})
     the Bellman function (\ref{model J}) is given by
        \beq
            \label{Bellman function answer power}
            J_P(w,x,v,t) = \frac{w^\gamma}{\gamma} f^{1/\delta} (v,t),
        \enq
     and  the optimal control is
        \beq
        \label{control answer power}
            \alpha_{P}^{*} (w, x, v, t) = \frac{w}{x (1-\gamma)} \left( \frac{\mu}{v} + \frac{\rho\,\sigma} {\delta} \frac{(\eta + \lambda + 1/2) }{v}\frac{M_{1+\lambda, \eta}\left(\Psi(v,t) \right)}{M_{\lambda, \eta} \left( \Psi(v,t) \right)}\right),
        \enq
        where
        \beqas
  \label{fvt answer power}
        f(v,t) & = & \frac{\Gamma (\eta - \lambda + 1/2)}{ \Gamma (2\eta+1)} e^{-\Psi(v,t)/2}
                \left( \Psi(v,t) \right) ^{\lambda} M_{\lambda, \eta} \left( \Psi(v,t)\right),\\
        \nonumber
\delta & = & 1+\rho^2\frac{\gamma }{1-\gamma},\quad
             \lambda = -\frac{k\Theta}{\sigma^2} +\frac{(1-\delta) \mu}{\rho \sigma}, \quad C= -\frac{\gamma}{1-\gamma} \frac{\mu^2}{2} \delta,\nonumber
            \quad \eta = \sqrt{\left( \lambda+\frac{1}{2}\right)^2 + \frac{2 C }{\sigma^2}},\\
        \Psi(v,t)& =& \frac{2kv}{\sigma^2 \left(e^{k(T-t)} - 1 \right)}, \nonumber
        \enqas
          $M_{\lambda, \eta}(z)$ is a Whittaker's function, and $\Gamma(z)$ is a gamma function.
        \end{thm}
          \begin{thm} For exponential utility (\ref{model U}) and process (\ref{model Heston})
       The Bellman function is given by
        \beq
        \label{Bellman function answer exp}
            J_E(x,w,v,t) = 1-\frac{e^{-c w}}{c} f^{1/{\delta}} (v,t),
        \enq
     and   the optimal control is
        \beq
            \alpha_{E}^{*} (w, x, v, t) = \frac{1}{c \,x} \left( \frac{\mu}{v} +
                \frac{\rho\,\sigma} {\delta}\frac{(\eta + \lambda + 1/2) }{v}\frac{M_{1+\lambda, \eta}\left(\Psi(v,t) \right)}{M_{\lambda, \eta} \left( \Psi(v,t) \right)}\right),
            \label{control answer exp}
        \enq
        where
        \beqas
  \label{fvt answer exp}
        f(v,t) & = &\frac{\Gamma (\eta - \lambda + 1/2)}{ \Gamma (2\eta+1)} e^{-\Psi(v,t)/2}
                \left( \Psi(v,t) \right) ^{\lambda} M_{\lambda, \eta} \left( \Psi(v,t)\right),\\
         \delta & = & 1-\rho^2,\quad \nonumber
             \lambda = -\frac{k\Theta}{\sigma^2} +\frac{(1-\delta) \mu}{\rho \sigma},\quad C= \frac{\mu^2}{2} \delta,
            \quad \eta = \sqrt{\left( \lambda+\frac{1}{2}\right)^2 + \frac{2 C }{\sigma^2}},\\
        \nonumber \Psi(v,t) &=& \frac{2kv}{\sigma^2 \left(e^{k(T-t)} - 1 \right)},
        \enqas
          $M_{\lambda, \eta}(z)$ is a Whittaker's function, and $\Gamma(z)$ is a gamma function.
    \end{thm}

{\it Comment.} The model (\ref{zar process}) with drift $\mu(X_t,t) = V_t$ is studied in \cite{Kraft}. For that particular model the coefficient for $f$ in (\ref{zar f}) is proportional to $v$. This is the reason why the affine representation is possible in that specific case. Moreover, if the boundary condition is not constant, the affine representation is not possible again. Our method, on the contrary,  allows us to obtain the required solution for any boundary conditions and for both models, either  Heston's or the model considered in \cite{Kraft}.

\section{Analysis}
\label{sec Analysis}

\subsection{Analysis of the optimal control}
Optimal control in (\ref{control answer power}) and (\ref{control answer exp}) is proportional to the sum of two terms:
the first term $\mu/v$ corresponds to the static portfolio optimization problem. It it simply the ratio of the instantaneous drift to the instantaneous variance and does not depend on time or on the parameters of the volatility process. The second term
\beq
\frac{\rho\,\sigma} {\delta} \frac{(\eta + \lambda + 1/2) }{v}\frac{M_{1+\lambda, \eta}\left(\Psi(v,t) \right)}{M_{\lambda, \eta} \left( \Psi(v,t) \right)}
\enq
represents the hedging of the future opportunity set ( see Merton \cite{Mer}). This term vanishes in a number of important special cases:
\begin{itemize}
\item[$\rho=0$:] no correlation between the two driving Brownian motions,
\item[$\gamma =0$:] when the investor has log-utility.
\end{itemize}

The case of zero volatility of volatility limit of Heston model ($\sigma$=0 in (\ref{model Heston})) turns the asset process into a process similar to a geometric Brownian motion but with a deterministically time varying volatility. If in addition we set $V_0=\Theta$ in (\ref{model Heston}), then we get a geometric Brownian motion.

Using the asymptotic results from Appendix \ref{appendix}  in case $\sigma \thicksim 0$, we get
\beq
f(v,t) \thicksim 1, \quad\quad\quad \frac{f_v(v,t)}{f(v,t)} \thicksim \frac{C}{k v^2} e^{k(T-t)}.
\enq
Therefore, for $\sigma \thicksim 0$ the Bellman function is approximately equal to the utility function, namely $J_P \thicksim U_P$, $J_E \thicksim U_E$; and the optimal controls are
\beqa
        \label{volvolzero answer power}
            \alpha_{P}^{*} (w, x, v, t) &=& \frac{w}{x (1-\gamma)} \left( \frac{\mu}{v} + \frac{\rho\,\sigma} {\delta_P} \frac{C_P}{k v^2} e^{k(T-t)}  \right),\\
  \alpha_{E}^{*} (w, x, v, t) &=& \frac{1}{c \,x} \left( \frac{\mu}{v} +
                \frac{\rho\,\sigma} {\delta_E}\frac{C_E}{k v^2} e^{k(T-t)}\right),
            \label{volvol zero answer exp}
        \enqa
        where constants $C_{[-]}$ and $\delta_{[-]}$  depend on the choice of utility.

\subsection{Hedging interpretation via a bond}
Suppose $C$ is given by (\ref{C power}) or (\ref{C exp}). With substitution $C/v = r$,  $f(v,t) = g(r,t)$ we can rewrite (\ref{heston f}) as
\beqa
\frac{b^2 r^3}{2}g_{rr} + \,h r \left( m  - r \right) g_r - r g + g_{t} =0, \nonumber \\
g(r,T) = 1, \nonumber
\enqa
where
\[
    b =\frac{\sigma}{\sqrt{C}} ,
    \qquad h = -\frac{\sigma^2}{C}(1+\lambda),
    \qquad m = -\frac{k C }{\sigma^2 (1+\lambda)}.
\]
Using the Feynman-Kac formula we get the following expression for $g=g(r,t)$
\beq
\label{3/2 bond}
g(r,t) =  \E e^{-\int_t^T r_s ds},
\enq
where process $r = (r_s)_{s \geq t}$ is a solution to the following stochastic differential equation
\beq
\label{3/2}
dr = h \, r (m - r) dt + b r^{3/2} dB^2, \quad\quad r_t =  C/v_t.
\enq
Equation (\ref{3/2}) describes the well-known $3/2$-model of stochastic interest rates (see \cite{Gatheral}.)
In this model, if $r$ is interpreted as the short rate, value of a zero coupon bond is given by (\ref{3/2}), see \cite{Brigo}.
Hence, the ratio $f_v /f$ can be expressed as
\beq
\label{hedging interpretation}
\frac{f_v(v,t)}{f(v,t)} = \frac{d}{dv} \log\E \left[ e^{-\int_t^T r_s ds} \,|\, r_t = C/v  \right].
\enq
Substituting (\ref{hedging interpretation}) into formulae for optimal control (\ref{control answer power}) and (\ref{control answer exp}) one can get an interpretation of the hedging of the future opportunity set via a bond.
\section{Appendix. Technical asymptotics relations}
\label{appendix}
      It follows from (\ref{answer f}), that function $f=f(v,t)$  can be written as a function of only $\Psi$, namely $f=f(\Psi)$, where $\Psi = \Psi(v,t)$ is given by (\ref{Psi}). This representation allows us to study the asymptotics of Bellman function and the optimal control by using the asymptotics of Whittaker functions $M_{\lambda, \mu}(z)$ (see, for example, \cite{Abramowitz}).

     Indeed, by using the well known formulae below
    \[
    M_{\lambda, \eta}(z) \thicksim z^{\eta+1/2}(1+O(z)),\quad z \rightarrow 0,
    \quad\quad\quad
    M_{\lambda, \eta}(z) \thicksim \frac{\Gamma(1+2\eta)}{\Gamma(1/2 - \lambda+\eta)} e^{z/2} z^{-\lambda}, \quad z \rightarrow \infty,
    \]
    we get
    \beq
    \label{asymt f 0}
    f(v,t) \thicksim \frac{\Gamma(1/2 - \lambda+\eta)}{\Gamma(1+2\eta)} \Psi ^{\eta + 1/2}(v,t)\left[ \Psi^{\lambda}(v,t) + O(\Psi(v,t))\right],\quad \Psi(v,t)  \rightarrow 0,
    \enq
    and
    \beq
     \quad\quad\quad f(v,t) \thicksim 1, \quad \Psi(v,t) \rightarrow \infty.
    \enq

    For $f_v(v,t)/f$ we have
    \beq
        \frac{f_v (v,t)}{f(v,t)} \thicksim \frac{(\eta+ \lambda +1/2) }{v} ,\quad \Psi(v,t) \rightarrow 0,
    \enq
    and
    \beq
      \frac{f_v (v,t)}{f(v,t)} \thicksim \frac{2 C}{\sigma^2} \frac{1}{v \Psi(v,t)},\quad \Psi(v,t) \rightarrow \infty.
    \enq

\smallskip
\noindent
{\bf Acknowledgements.} The authors are grateful to Yury Kabanov for fruitful discussions and useful suggestions.

\end{document}